\documentclass[a4paper]{article}

\usepackage{INTERSPEECH2022}
\usepackage{booktabs, multirow}
\usepackage{hyperref}
\usepackage{amsmath}
\usepackage{float}

\title{Speaker Recognition in the Wild}
\name{Neeraj Chhimwal$^1$, Anirudh Gupta$^1$, Rishabh Gaur$^1$, Harveen Singh Chadha$^1$, Priyanshi Shah$^1$, Ankur Dhuriya$^1$, Vivek Raghavan$^2$}
\address{
	$^1$Thoughtworks\\
	$^2$Ekstep Foundation}
\email{neeraj.chhimwal@thoughtworks.com, vivek@ekstep.org}
\email{\{neeraj.chhimwal, anirudh.gupta, rishabh.gaur, harveen.chadha, priyanshi.shah, ankur.dhuriya  \}@thougtworks.com, vivek@ekstep.org}
\begin{document}

\maketitle
\begin{abstract}
In this paper, we propose a pipeline to find the number of speakers, as well as audios belonging to each of these now identified speakers in a source of audio data where number of speakers or speaker labels are not known a priori. We used this approach as a part of our Data Preparation pipeline for Speech Recognition in Indic Languages. \footnote{\url{https://github.com/Open-Speech-EkStep/vakyansh-wav2vec2-experimentation}}

To understand and evaluate the accuracy of our proposed pipeline, we introduce two metrics- Cluster Purity, and Cluster Uniqueness. Cluster Purity quantifies how "pure" a cluster is. Cluster Uniqueness, on the other hand, quantifies what percentage of clusters belong only to a single dominant speaker. We discuss more on these metrics in section \ref{sec:metrics}.

Since we develop this utility to aid us in identifying data based on speaker IDs before training an Automatic Speech Recognition (ASR) model, and since most of this data takes considerable effort to scrape, we also conclude that 98\% of data gets mapped to the top 80\% of clusters (computed by removing any clusters with less than a fixed number of utterances- we do this to get rid of some very small clusters and use this threshold as 30), in the test set chosen.
\end{abstract}

\section{Introduction}
Given a source of audio data, with no prior knowledge of number of speakers or speaker labels, our goal is to find the number of speakers and a mapping of these identified speaker labels to audio utterances in a corpus. In a supervised setting, where the goal is to either recognise or verify a speaker among a group of pre-enrolled speakers, is also an active area of research where deep learning techniques have become the new state-of-the-art for these tasks \cite{sztaho2019deep} . Common tasks in supervised approach can be grouped into two major categories- Speaker Recognition and Speaker Verification. Speaker Recognition is the identification of a person from characteristics of voices and is used to answer the question "Who is speaking?". Speaker verification is the verification of a speaker’s claim of their identity.
It is used to answer the question "Is this really the mentioned speaker speaking?" However, in an unsupervised setting, which is what we're dealing with in our use case, both of the above mentioned tasks aren't possible. 

In our case, the data is scraped from the web with almost no metadata about speakers. We’re talking about unlabeled audios belonging to unknown, previously unidentified speakers. We’ll still be solving a speaker recognition task, but completely unsupervised. This is formally  known as Speaker Clustering. Speaker clustering is the task of identifying the unique speakers in a set of audio recordings without knowing who and how many speakers are present in the entire data. We assume that each of these audio recordings (referred to as utterance in the paper) belongs to exactly one speaker. We achieve this by using Voice Activity Detection on the full audios to break them into smaller chunks or utterances, as a part of our data preparation steps. This ensures that the audio is short enough to accommodate only one unique speaker, as it is cut from one silence to the next. 
We use an open-sourced pre-trained Neural Network \footnote{\url{https://github.com/resemble-ai/Resemblyzer}} as an embedding technique and generate deep embeddings for every utterance. These embeddings are then fed to our clustering stage where we primarily use HDBSCAN \cite{campello2013density} with some nuances to tackle the large number of tunable hyper-parameters- this is done to reduce effort in finding optimal parameters for every new source of audio data we scrape. All the steps involved will be explained in detail in section \ref{sec:pipeline}.

\section{Pipeline}\label{sec:pipeline}
\subsection{Deep Embeddings}
We use an open-sourced pre-trained Neural Network as an embedding technique\footnotemark[2]. Which means that given an audio file of speech, this network creates a summary vector of 256 values (also known as "embedding") that summarizes the characteristics of the voice spoken. This model is speaker-discriminative and has been trained on a very particular text-independent speaker verification task, trained to optimize Generalized End to End loss for Speaker Verification \cite{wan2018generalized}.

The inputs to this model are 40-channels log-mel spectrograms with a 25ms window width and a 10ms step. The output is the L2-normalized hidden state of the last layer, which is a vector of 256 elements. 

\begin{figure}[h]
	\centering
	\includegraphics[width=1\linewidth]{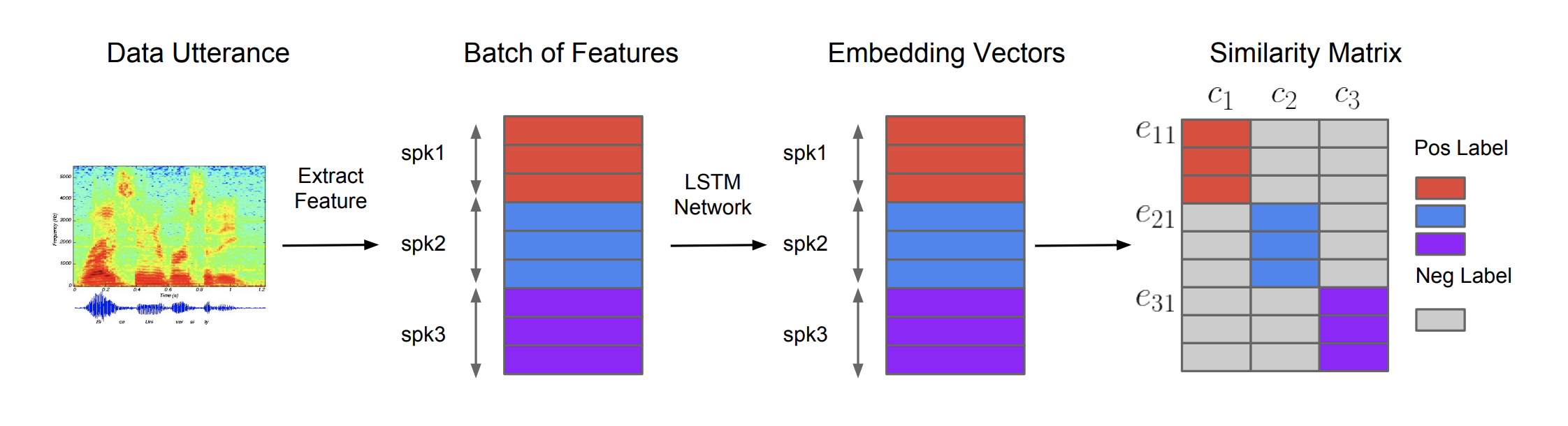}
	\caption{System overview of GE2E extracted from (Wan et al., 2017). Different colors indicate utterances/embeddings from different speakers.}
	\label{fig:g2e2}
\end{figure}

However, the data used to train this model belongs to only one language- English, while our corpus belongs to Indic languages. This model was trained using more than 1000 hours of English data from the sources: LibriSpeech-other, VoxCeleb1, and VoxCeleb2, and contains audios belonging to 1.8 thousand speakers. To ascertain that the embeddings are also able to encode speaker information for Hindi, we did the following experiment:   
We used Resemblyzer's trained voice encoder model to generate embeddings for all audios in our test dataset comprising of 20 hours of audios belonging to 80 speakers. On plotting a distribution plot of the cosine similarities between embeddings belonging to same speakers, and those belonging to different speakers, it's clear that the model trained on English data is still able to encode speaker information for Hindi.

\begin{figure}[h]
	\centering
	\includegraphics[width=1\linewidth]{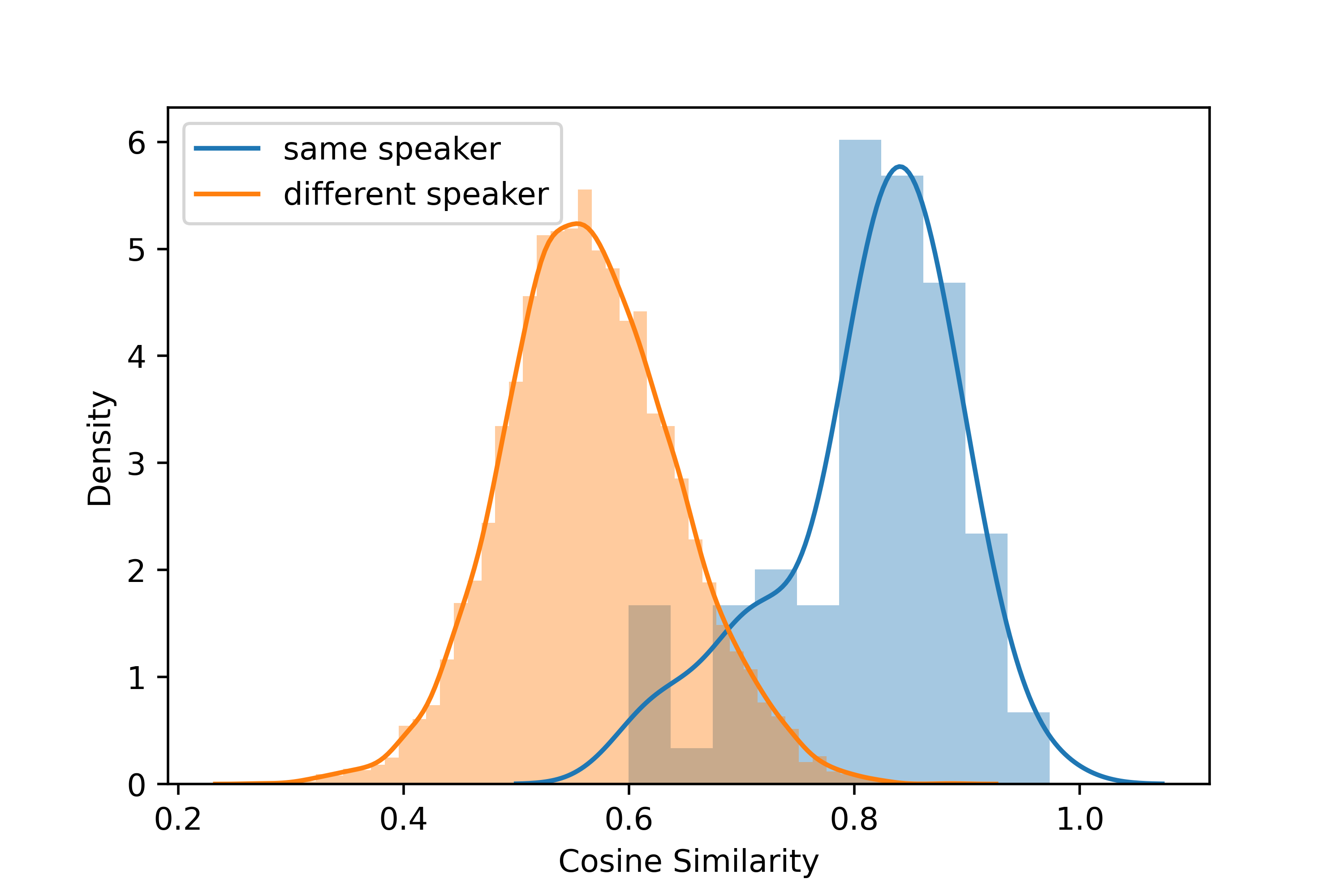}
	\caption{Cosine Similarity between same vs different speaker embeddings}
	\label{fig:sp_sep}
\end{figure}

\subsection{Clustering algorithm}
We use Hierarchical Density-Based Spatial Clustering of Applications with Noise \cite{campello2013density} (HDBSCAN) as the main clustering algorithm. 

HDBSCAN supports a special metric called "precomputed". So if we create the clusterer with the metric set to precomputed then the clusterer will assume that, rather than being handed a vector of points in a vector space, it is receiving an all pairs distance matrix. This is the approach we use. So essentially, instead of handling a 256 dimensional vector for each utterance, we calculate pairwise cosine distances between each embeddings and use this pre-computed distance matrix (which is a square matrix) as our input to the clusterer.

We found that for sources with more than 15-20 hours of data, this was not ideal as the clustering algorithm runs out of memory while trying to fit these data points. Another issue was that some "big" clusters contained majority of data, whereas some clusters, although belonging to the same original speaker were present as different "small" clusters. So we devised a strategy to tackle all of these issues.

\begin{enumerate}
 \item Since many of our sources contained data in hundreds of hours, we decided to implement a partial set strategy where clusters will be computed for this partial set (roughly corresponding to 20 hours of data at once, this number can be customized).

 \item Once all the clusters from these partial sets have been found, we apply repetitive merging of clusters over a range of cosine similarities with decay (default merging starts from 96\% and goes down to 90\% at 1\% intervals to accommodate for clusters that can be merged after every iteration), where smaller homogeneous clusters belonging to the same speaker but from same/different partial sets can merge to form bigger clusters for single speakers. The way cosine similarities work in this case is, we compute a mean cluster embedding for each cluster, and then compute similarities between each cluster as if it was just one centroid point of 256 dimensions.

 \item After this step, we compute mean cluster size and define a method to identify "big" clusters based on variance from this mean cluster size. These clusters, in our experiments were usually the ones with a large number of speakers grouped as one (could be some speakers belonging to the same gender). But this was not always the case, as sometimes some speakers actually had a lot of data, and we would not like to break these well formed groups. To tackle this issue, we run clustering again on these big clusters but with a slight difference: we change the cluster selection method, which determines how flat clusters are selected from the cluster tree hierarchy. The default method is 'eom' for Excess of Mass, which is not always the most desirable approach to cluster selection. In cases where we are more interested in having small homogeneous clusters, we may find that Excess of Mass has a tendency to pick one or two large clusters and then a number of small extra clusters. We choose 'leaf' as a cluster selection method while splitting these big clusters, to allow the algorithm to select leaf nodes from the tree, producing many small homogeneous clusters. In our experiments, we verify that big clusters with just one speaker were not split using this method, which allows some mistakes in identifying non-homogeneous big clusters since big homogeneous clusters seem to be immune to splitting.

 \item After step 3, we have a list of clusters that were unchanged plus the new list of leaf clusters. At this step, we again do the same repetitive merging to let these new smaller clusters combine to form better groupings for individual speakers.

 \item Since HDBSCAN also takes into account the "noise", meaning not every point is allocated to some cluster. At every clustering step, some points get classified as noise and we keep a track of these points. In this step, we allow these noise points to be merged with the clusters they are closest to, as long as this similarity is greater than the "fit noise point on similarity" parameter. This "noise" is not to be confused with environmental noise or any audio specific noise we may encounter, this strictly refers to data points that could not be fit into any cluster.
\end{enumerate}

\section{Hyperparameters}
We keep these parameters fixed in order to automate our pipeline with many moving parts (such as gender identification and language identification for each utterance). But you can play around if you know what kind of data you are expecting in terms of a rough idea about average speaker duration, etc.

\begin{enumerate}
\item partial\_set\_size (default=10,000): if the number of utterances in the source is greater than partial\_set\_size, clustering is done over these partial sets for step 1. We use 10,000 as it roughly corresponds to 20 hours of data for us. 

  \item min\_cluster\_size (default=4): this is the smallest size grouping that you wish to consider a cluster. 
  \item min\_samples (default=1): this is the number of points required in the neighbourhood of a point to be considered a core point. min\_samples value can range from 1 to min\_cluster\_size. Smaller values of min\_samples have an effect of lowering the number of points classified as noise.
   \item metric (default='precomputed'): we use 'precomputed' since we and pass a distance matrix as input to HDBSCAN.
  \item fit\_noise\_on\_similarity (default=0.8): since all the clustering steps will lead to some points not being mapped to any cluster, we try to fit these in clusters with a similarity score higher than the value of fit\_noise\_on\_similarity, in order to reduce data loss.
\end{enumerate}

\section{Metrics proposed}\label{sec:metrics}
We propose two metrics to evaluate the clusters formed which can help you in getting a sense of the speaker clusters identified. 

\subsection{Cluster Purity}
Cluster Purity quantifies how "pure" a cluster is. We define Cluster Purity as the ratio of number of audio utterances belonging to the dominant speaker in a cluster to the total number of audio utterances present in the same cluster. 

For every speaker cluster:

\begin{equation}
CP  = {{num\ utterances\ belonging\ to\ dominant\ speaker} \over {total\ utterances\ in\ the\ cluster} }
\end{equation}

\subsection{Cluster Uniqueness}
Cluster Uniqueness quantifies what percentage of clusters belong only to single dominant speakers. In our experiments, we identified that some clusters even when belonging to the same speaker ID, were not able to merge because these two clusters had a low cosine similarity score. This may happen if there is noise in the background for some audios or if the audio isn't representative enough (has less content). Since some speakers may be present in many clusters at once, this metric helps in identifying what percentage of the original true speakers were grouped in their respective clusters only.

\begin{equation}
CU  = {{num\ speakers\ with\ only\ one\ dominant\ cluster} \over {total\ number\ of\ clusters} }
\end{equation}

\section{Results}
Below are the results on our private test set of 80 speakers containing 20 hours of audio data. This dataset has a 50-50 gender split, meaning 40 voices are male and 40 are female. On running any source through our clustering, there are some very small clusters that we manually remove. For this source, the number of clusters identified was 104, but on removing clusters with less than 30 utterances (average length of an utterance is 6 seconds), we get cluster count as 79. This is one part where manual effort can be used to find a better threshold for minimum cluster size allowed, depending on the task. 

\begin{table}[H]
	\centering
	\caption{Results on test set}
	\label{tab:results}
	\scriptsize
	\begin{tabular}{lrrr}\toprule
		\textbf{Metric} &\textbf{Result} \\\midrule
		\textbf{num clusters identified} &79\\
		\textbf{average cluster purity} &96.00\ \%\\
		\textbf{num speakers present in only one cluster} &67\\
		\textbf{cluster uniqueness} &84.81\ \%\\
		\textbf{utterances classified as noise} &1.35\ \%\\
		\bottomrule
	\end{tabular}
\end{table}

\section{Discussion}
In this paper, we have identified a way to leverage an open-sourced speaker-discriminative and text-independent Speaker Verification based neural network, trained for English, as a Voice Encoder to use for the downstream task of clustering Hindi audios. But we can run into many issues with the clusters formed, depending on the audio quality of the source- as this is something we didn't explore. Our use case allowed us to cap the per-speaker audio data at 90 minutes per speaker identified. In our experiments, we found that this approach worked well for most of the sources, meaning all the components were mostly working as hypothesized. 

\section{Conclusion \& Future work}
In this direction, an area that can be directly explored is the effect on speaker separation when all the speakers in a source belong to one gender only- male or female voices. This is important since in our experiments, some clusters contained two female speakers with similar voices. This similarity can be explored further to mitigate this effect. 
In this approach, we derive our input from a deep neural network and then use a classical unsupervised clustering algorithm with added nuances. Our clustering algorithm is model-free and works on a pre-defined distance measure that we provide as a distance matrix between our high dimensional embeddings. Future work can explore both the voice encoder training step for Indic languages as well as deep clustering approaches to minimize the dependency on clusterer related hyper parameters. A detailed look into the various classical clustering algorithms \cite{li2021discriminative, murtagh2014ward, von2007tutorial, verma2003comparison, shum2013unsupervised} and loss functions \cite{wan2018generalized, hershey2016deep, le2018robust}, will also help in identifying the best approach possible, with respect to compute and time constraints as well as accuracy.

\section{Acknowledgment}
All authors gratefully acknowledge Ekstep Foundation for supporting this project financially and providing infrastructure. A special thanks to Dr. Vivek Raghavan for constant support, guidance and fruitful discussions. We also thank Nikita Tiwari, Ankit Katiyar, Heera Ballabh, Niresh Kumar R, Sreejith V, Soujyo Sen, Amulya Ahuja and Rajat Singhal for helping out when needed and extending infrastructure support for data processing and model training.

\bibliographystyle{IEEEtran}

\bibliography{mybib}

\end{document}